%% file: paper_v27.tex
\RequirePackage{lineno}
\documentclass[prl,twocolumn,showpacs,amsmath,amssymb,superscriptaddress,nofootinbib]{revtex4-1}

\usepackage{epsfig}
\usepackage{graphicx}
\usepackage{dcolumn}
\usepackage{bm}
\usepackage{color}
\usepackage{url}
\usepackage{multirow}
\usepackage{enumitem}

\newcommand {\snn}	{\sqrt{s_{_{\rm NN}}}}

\newcommand {\pA}	{$p$+Au}
\newcommand {\dA}	{$d$+Au}

\newcommand {\AuAu}	{Au+Au}
\newcommand {\pPb}	{$p$+Pb}
\newcommand {\PbPb}	{Pb+Pb}
\newcommand {\Bvec}	{\vec{B}}

\newcommand {\phia}	{\phi_{\alpha}}
\newcommand {\phib}	{\phi_{\beta}}
\newcommand {\phic}	{\phi_c}
\newcommand {\pt}	{p_{T}}
\newcommand {\vc}	{v_{2,c}}
\newcommand {\vtwo}	{v_2\{2\}}
\newcommand {\vres}	{v_{2,{\rm res.}}}
\newcommand {\gOS}	{\gamma_{\rm OS}}
\newcommand {\gSS}	{\gamma_{\rm SS}}
\newcommand {\dg}	{\Delta\gamma}
\newcommand {\dgscale}	{\dg_{\rm scaled}}

\newcommand {\phires}	{\phi_{\rm res.}}

\newcommand {\dNch}	{dN_{\rm ch}/d\eta}

\newcommand {\mean}[1]	{\langle #1\rangle}
\newcommand {\note}[1]	{}

\begin{document}
\title{Charge-dependent pair correlations relative to a third particle in $p$+Au and $d$+Au collisions at RHIC}
\input{star_authors.tex}

\date{\today}

\begin{abstract}
Quark interactions with topological gluon configurations can induce chirality imbalance and local parity violation in quantum chromodynamics. 
This can lead to electric charge separation along the strong magnetic field in relativistic heavy-ion collisions -- the chiral magnetic effect (CME). 
We report measurements by the STAR collaboration of a CME-sensitive observable in $p$+Au and $d$+Au collisions at 200 GeV, 
where the CME is not expected, using charge-dependent pair correlations relative to a third particle. 
We observe strong charge-dependent correlations similar to those measured in heavy-ion collisions. 
This bears important implications for the interpretation of the heavy-ion data.
\end{abstract}
\pacs{25.75.-q, 25.75.Gz, 25.75.Ld}
\maketitle


\section{Introduction}
In quantum chromodynamics, interactions of massless quarks with fluctuating topological gluon fields are predicted to induce chirality imbalance and parity violation in a local domain ~\cite{Lee:1974ma,Kharzeev:1998kz,Kharzeev:1999cz}.
This chirality imbalance can lead to an electric charge separation in the presence of a strong magnetic field ($\Bvec$), 
a phenomenon known as the chiral magnetic effect (CME)~\cite{Fukushima:2008xe,Muller:2010jd,Liu:2011ys,Kharzeev:2004ey,Kharzeev:2007jp,Kharzeev:2013ffa}. 
Such a strong $\Bvec$-field may be available in relativistic heavy-ion collisions, generated by the incoming protons at early times~\cite{Kharzeev:2007jp,Asakawa:2010bu}. 
Extensive theoretical and experimental efforts have been devoted to the search for the CME-induced charge separation along $\Bvec$ in heavy-ion collisions~\cite{Kharzeev:2015znc,Zhao:2018ixy,zhao:225}. 

The commonly used observable to search for charge separation in heavy-ion collisions is the three-point azimuthal correlator~\cite{Voloshin:2004vk},
\begin{linenomath}
\begin{equation}
	\gamma\equiv\cos(\phia+\phib-2\psi),
	\label{eq:1p}
\end{equation}
\end{linenomath}
where $\phia$ and $\phib$ are the azimuthal angles of particles $\alpha$ and $\beta$, respectively.
In Eq.~(\ref{eq:1p}), $\psi$ is the azimuthal angle of the impact parameter vector. 
In heavy-ion collisions, it is called the reaction plane (spanned by the impact parameter direction and the beam). 
It is often approximated by the second order harmonic participant plane ($\psi_2$)~\cite{Poskanzer:1998yz,Alver:2006wh},
constructed experimentally by the event plane measured from final state particle azimuthal distribution. 
To measure the $\gamma$, instead of using the event plane, the three-particle correlator method is often used~\cite{Voloshin:2004vk,Abelev:2009ad,Abelev:2009ac}:
\begin{linenomath}
\begin{equation}
   \gamma=\mean{\cos(\phia+\phib-2\phic)}/\vc\,,
   \label{eq:3p}
\end{equation}
\end{linenomath}
where $\phic$ is the azimuthal angle of a third, charge-inclusive particle $c$ which serves as a measure of the $\psi$.
The imprecision in determining the $\psi$ by a single particle is corrected by a resolution factor,
equal to the second-order Fourier coefficient of particle $c$'s azimuthal distribution, $\vc$, 
also known as the elliptic flow~\cite{Reisdorf:1997fx}.  
In order to remove the charge independent background~\cite{Abelev:2009ad,Abelev:2009ac}, 
such as that due to momentum conservation, 
the correlation difference variable is used, 
\begin{linenomath}
\begin{equation}
\dg\equiv\gOS-\gSS, 
\label{eq:4p}
\end{equation}
\end{linenomath}
where $\gOS$ stands for the correlation of opposite-sign (OS) pairs ($\alpha$ and $\beta$ have opposite-sign electric charges) and
$\gSS$ for that of the same-sign (SS) pairs ($\alpha$ and $\beta$ have same-sign electric charge).



Significant $\dg$ is indeed observed in heavy-ion collisions at RHIC~\cite{Abelev:2009ad,Abelev:2009ac,Adamczyk:2013hsi,Adamczyk:2014mzf}, and at LHC~\cite{Abelev:2012pa,Acharya:2017fau,Khachatryan:2016got,Sirunyan:2017quh}.
However, a decisive answer regarding the existence, or not, of the CME is still under debate. 
The main difficulty in interpreting the $\dg$ observable as originated from the CME is the possibility of significant charge-dependent background contributions, 
such as those from resonance decays~\cite{Voloshin:2004vk,Wang:2009kd,Bzdak:2009fc,Schlichting:2010qia,Adamczyk:2013kcb,Wang:2016iov}.  
This is because the $\dg$ variable is ambiguous between an OS pair from the CME back-to-back perpendicular to $\psi_2$ 
(charge separation) and an OS pair from a resonance decay along $\psi_2$ (charge conservation). 
There are more particles/resonances along the $\psi_2$ (or the particle $c$) direction than perpendicular to it, 
an effect quantified by the elliptical anisotropy parameter $\vres$. 
Equation~(\ref{eq:3p}) is valid and $\dg$ would be a good measure of the CME only under the assumption that all particles 
(including the CME-related particles) are correlated to a global plane $\psi_2$, but intrinsically uncorrelated among themselves. 
When $\alpha$ and $\beta$ are intrinsically correlated, then $\dg$ would contain a background ($\dg_{\rm bkgd}$), 
arising from the coupling of this elliptical anisotropy and the intrinsic decay correlation and is expected to take the following form~\cite{Voloshin:2004vk,Wang:2009kd,Wang:2016iov}:
\begin{linenomath}
\begin{equation}
	\dg_{\rm bkgd}\propto\mean{\cos(\phia+\phib-2\phires)}\vres\,.
	\label{eq:bkgd}
\end{equation}
\end{linenomath}
Other possible backgrounds include three-particle nonflow correlations, where the correlation of particle $\alpha$, $\beta$ with particle $c$ is also of nonflow nature. 
Moreover, the estimate of $\vc$ via two-particle correlations may also be affected by short-range nonflow correlations. 
These effects are likely dominant for very low multiplicity events because they are not sufficiently diluted by multiplicity combinatorics.
Nevertheless, the factorization relation in Eq.~(\ref{eq:3p}) is still expected to approximately hold, 
regardless of the nature of the background correlations~\cite{Kikola:2011tu}.


In non-central heavy-ion collisions, the participant plane, although fluctuating~\cite{Alver:2006wh}, 
is generally aligned with the reaction plane, 
thus generally perpendicular to $\Bvec$. 
In proton-nucleus collisions, however, the participant plane is determined purely by geometry fluctuations, 
and thus is essentially uncorrelated with the impact parameter or the $\Bvec$ direction~\cite{Khachatryan:2016got,Belmont:2016oqp,Kharzeev:2017uym}. 
A recent study, considering the fluctuating size of the proton, suggests a small but non-zero correlation~\cite{Kharzeev:2017uym}. 
Therefore, CME-induced $\dg$ with respect to the $\psi_{2}$ is significantly suppressed in proton-nucleus collisions compared to possible signals from heavy-ion collisions~\cite{Kharzeev:2017uym}.
Background correlations aforementioned is expected to be present in proton-nucleus collisions as well.
These correlations are propagated to the three-particle correlator via correlations with respect to particle $c$,
not directly to the impact parameter or the $\Bvec$ direction.
Thus, the backgrounds in proton-nucleus collisions contribute in a similar fashion as those in heavy-ion collisions.
Indeed, a large $\dg$ signal was observed in \pPb\ collisions at the LHC, similar to that in \PbPb\ collisions. 
This challenged the CME interpretation of the heavy-ion data~\cite{Khachatryan:2016got}. 


It is possible that the CME would decrease as collision energy increases,
due to the more rapidly decaying $\Bvec$ at higher energies~\cite{Kharzeev:2007jp,Skokov:2009qp}. 
Hence, the similarity between \pPb\ and \PbPb\ collisions at $\sqrt{s_{\rm NN}}$ = 5.02 TeV at the LHC may be expected, and the situation at RHIC could be different~\cite{Kharzeev:2015znc}. 
Here we report $\dg$ measurements by the STAR experiment at RHIC in small-system \pA\ and \dA\ collisions at $\sqrt{s_{\rm NN}}$ = 200 GeV.

\section{Experiment and data analysis}
The data reported here were taken by the STAR experiment in 2003 (\dA) and 2015 (\pA). The STAR experiment apparatus is described elsewhere~\cite{Ackermann:2002ad}. 
Minimum bias (MB) triggers were used for both data taking periods. 
For \dA~\cite{Adams:2003im}, the MB trigger required at least one beam-rapidity neutron in the Zero
Degree Calorimeter (ZDC)~\cite{Adler:2003sp} in the Au beam direction.
For \pA, the MB trigger data used in this analysis was defined as a coincidence between the two Vertex Position Detectors (VPDs)~\cite{Llope:2003ti}.

The detectors relevant to this analysis are the cylindrical Time Projection Chamber (TPC)~\cite{Anderson:2003ur,Ackermann:1999kc} 
residing inside an approximately uniform magnetic field of 0.5 Tesla along the beam direction ($z$). 
Charged particles traversing the chamber ionize the TPC gas. 
The ionization electrons drift towards the TPC endcaps in a uniform electric field, 
provided by the high voltage on the TPC central membrane. The avalanche electrons are collected by the pad planes, 
and together with the drift time information, 
provide three-dimensional space points of the ionization called ``hits''.

Trajectories are reconstructed from those hits; at least 10 hits are required for a valid track.
The interaction's primary vertex is reconstructed from charged particle tracks. 
Tracks with the distance of closest approach (DCA) to the primary vertex within 3~cm are considered primary tracks. 
The data are reported as a function of the efficiency corrected charged particle multiplicity density $\dNch$ at mid-rapidity~\cite{Abelev:2008ab}. 
The efficiency is estimated via the STAR standard embedding procedure, which is $\sim$ 93$\%$ in \pA\ and \dA\ collisions.

In this analysis, events with primary vertices within 30~cm in \pA\ (50~cm in \dA) longitudinally and 2~cm in \pA\ (3.5~cm in \dA) transversaly from the geometrical center of the TPC are used. 
To ensure high quality of primary particles, 
further selections are applied to require tracks with at least 20 hits and DCA less than 2~cm. 
Split tracks are removed by requiring the
number of hits over the maximum number of possible hits to be greater than 0.52~\cite{Adamczyk:2015lme}.
In the \pA\ analysis, where VPD detectors and Time-of-Filght (TOF) detector~\cite{STARtof:111} 
are available, the primary vertex is required to match with the VPD's measured vertex within 6~cm,
and primary tracks are required to match with the TOF detector in order to reduce the pile-up tracks. 

Tracks in the full TPC acceptance ($|\eta|<1$, reducing to $|\eta|<$~0.9 in case of TOF matched tracks in \pA) 
with transverse momentum $\pt$ from 0.2 to 2.0 GeV/$c^{2}$
are used for all three particles in the three-particle correlator of Eq.~(\ref{eq:3p}). 
The cumulant method is used to compute $\gamma$, where the calculation loops over the $\alpha$ and $\beta$ particles, 
and the particle $c$ is handled by the cumulant of the remaining particles except $\alpha$ and $\beta$. 
No $\eta$ gap is applied between any pair among the three particles as in Refs.~\cite{Abelev:2009ad,Abelev:2009ac}. 
The $\vc$ is obtained by the two-particle cumulant~\cite{Bilandzic:2010jr}. To gauge the nonflow effects, various $\eta$ gaps of 0, 0.5, 1.0 and 1.4 are applied.
The $p_{T}$-dependent TPC tracking efficiency is not corrected for the $\gamma$ correlator as in Refs.~\cite{Abelev:2009ad,Abelev:2009ac}, 
and this effect is included in the systematic uncertainties. 
The detector non-uniform azimuthal acceptance effect is corrected by the recentering method as a function of $p_{T}$~\cite{Adamczyk:2013gw,Selyuzhenkov:2007zi}.


\section{Systematic Uncertainties}
The systematic uncertainties are estimated as follows.
The required minimum number of points is varied from 20 to 25. 
The DCA of tracks is varied from 2 cm to 1 and 3 cm. 
The $p_{T}$ range of the particle $c$  is varied from 0.2-2 GeV/$c$ to 0.2-5 GeV/$c$.
The difference between the results from events with positive and negative reconstructed $z$ coordinate of primary vertex is $\sim 2\%$.
The $p_{T}$-dependent TPC tracking efficiency correction introduces a $\sim 1\%$ difference. 
$p_{T}$-independent azimuthal non-uniformity recentering correction is also studied.
The TOF detector acceptance is limited to $|\eta| < 0.9$, and this causes a $\sim -6\%$ (single sided) effect in \pA.  
The systematic uncertainties obtained by various cuts and sources are added in quadrature. 
These are plotted in the figures as brackets. 
The horizontal brackets indicate the systematic uncertainty of the $\dNch$.
The vertical brackets indicate the systematic uncertainty of the correlator. 
Total systematic uncertainty of the $\dg$ is $\sim 9\%$ in \pA\ and in \dA\ (Table~\ref{systable}). 
Total systematic uncertainty of the $\dNch$ is $\sim 15\%$ in \pA\ and is $\sim 7\%$ in \dA.

\begin{table}[!htpb]
\centering
\begin{tabular}{l|c|r}
    \hline
    \hline
    source &  \pA\ & \dA\ \\
    \hline
	dca \& nHits   & $\pm 5\%$  & $\pm 8\%$  \\ 
	$p_{T}$(c)     & $\pm 0\%$  & $\pm 1\%$  \\
    $V_{z}$                             & $\pm 2\%$  & $\pm 2\%$ \\
	$p_{T}$-dependent efficiency        & $\pm 1\%$  & $\pm 1\%$ \\
	$p_{T}$-independent non-uniformity  & $\pm 5\%$  & $\pm 4\%$ \\
	TOF acceptance                      & $- 6\%$  &  -- \\
    \hline
	total  &  $^{+7}_{-9}\%$   & $\pm 9\%$ \\
\hline
\hline
\end{tabular}
\caption{The systematic uncertainties of the $\dg$ correlator in \pA\ and in \dA\ collisions.}
\label{systable}
\end{table}



\section{Results and discussions}
\begin{figure}[hbt]
  \begin{center}
    \includegraphics[width=0.45\textwidth]{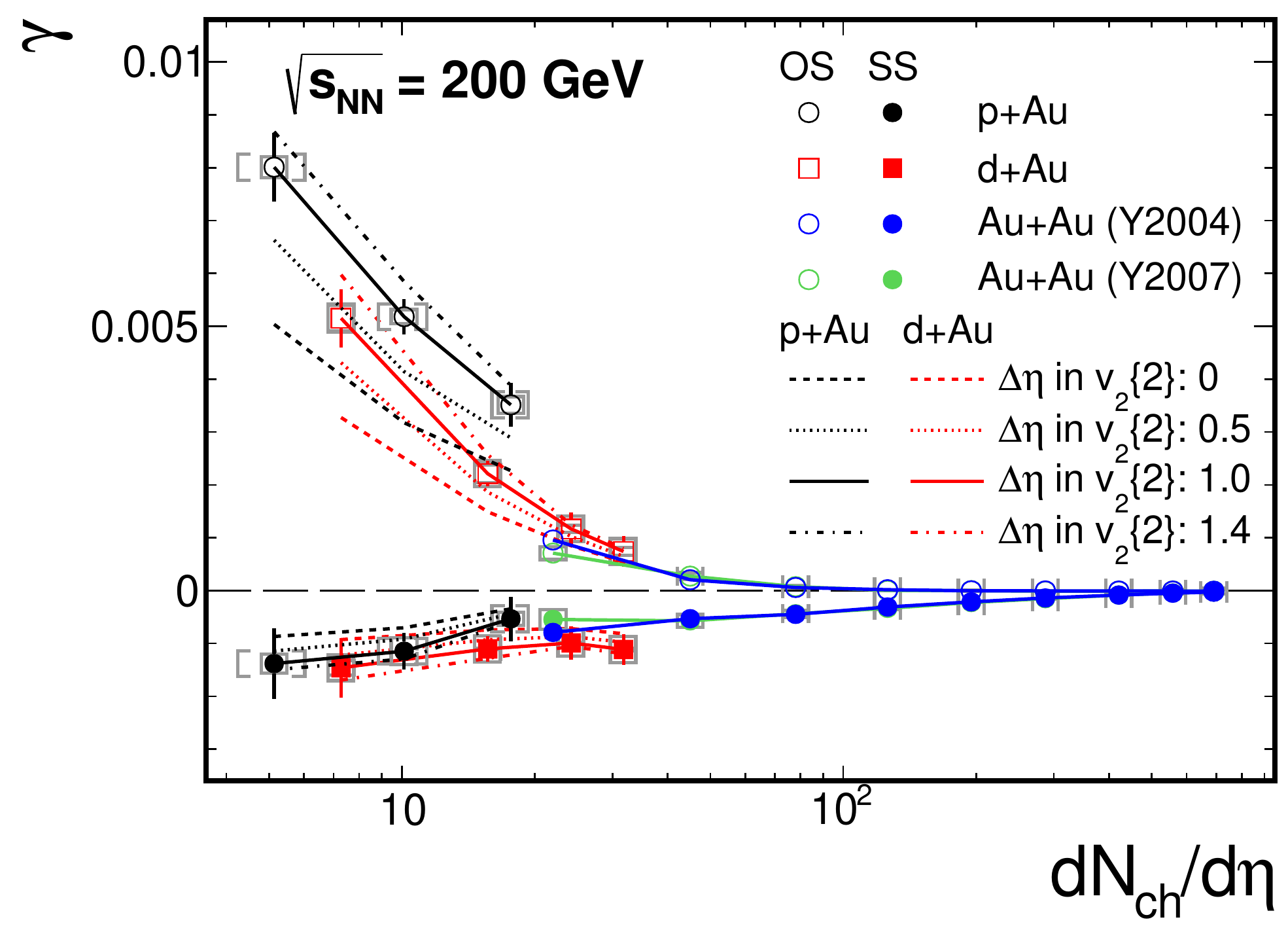}
	  \caption{The $\gSS$ and $\gOS$ correlators in \pA\ and \dA\ collisions as a function of multiplicity, compared to those in \AuAu\ collisions~\cite{Abelev:2009ad,Abelev:2009ac,Adamczyk:2013hsi}. Particles $\alpha, \beta,$ and $c$ are all from the full TPC $|\eta|<1$; no $\eta$ gap is applied. The $\vc$ is obtained by two-particle cumulants with $\eta$ gap of 1.0; 
	  results with $\eta$ gaps of 0, 0.5 and 1.4 are shown as dashed lines.	
	  Statistical errors are shown by the vertical bars and systematic uncertainties are shown by the vertical brackets.
	  The horizontal brackets indicate the systematic uncertainty of the $\dNch$. 
	  }
    \label{fig:g}
  \end{center}
\end{figure}

Figure~\ref{fig:g} shows the $\gSS$ and $\gOS$ results as functions of multiplicity in \pA\ and \dA\ collisions at $\snn=200$~GeV. 
For comparison, the corresponding \AuAu\ results~\cite{Abelev:2009ad,Abelev:2009ac,Adamczyk:2013hsi} are also shown. 
The dashed lines represent the results with $\vc$ using different $\eta$ gaps of 0, 0.5 and 1.4 in \pA\ and \dA\ collisions.
The results with $\vc$ using $\eta$ gaps of 1.0 in \pA\ and \dA\ collisions are plotted as solid lines.
The results show that the variation from different $\eta$ gaps is large but tends to converge towards high multiplicity.
The $\gSS$ and $\gOS$ results seem to follow a 
decreasing trend with increasing multiplicity in all systems.

\begin{figure}[hbt]
  \begin{center}
    \includegraphics[width=0.45\textwidth]{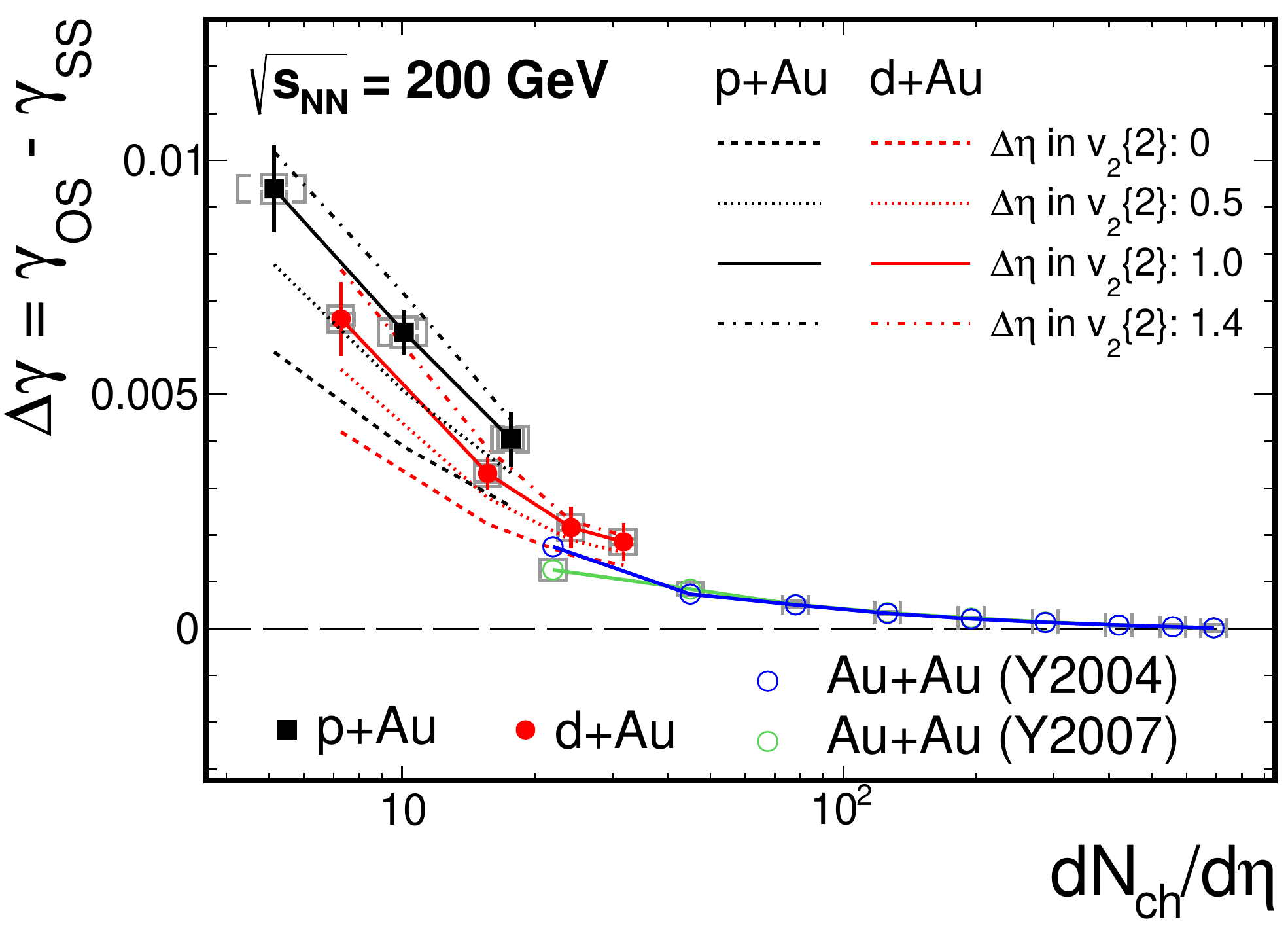}
    \caption{The $\dg$ correlator in \pA\ and \dA\ collisions as a function of multiplicity, compared to that in \AuAu\ collisions~\cite{Abelev:2009ad,Abelev:2009ac,Adamczyk:2013hsi}. The difference measures the charge-dependent correlations. 
	  The data points connected by solid lines are measured using $\Delta\eta$ gap of 1.0 in $\vtwo$. 
	  Dashed lines represent the results using $\vc$ with $\eta$ gaps of 0, 0.5 and 1.4.}
    \label{fig:dg}
  \end{center}
\end{figure}

Figure~\ref{fig:dg} shows $\dg$ as a function of multiplicity in \pA\ and \dA\ collisions, and, for comparison, 
in \AuAu\ collisions~\cite{Abelev:2009ad,Abelev:2009ac,Adamczyk:2013hsi}. The $\dg$ decreases with increasing multiplicity in both systems. 
Large $\dg$ values are observed in \pA\ and \dA\ collisions, comparable to the peripheral \AuAu\ collision data at similar multiplicities. 
Our new $p$+Au and $d$+Au measurements demonstrate that background contributions could produce magnitudes of the $\dg$ correlator comparable to what has been observed in Au+Au data, 
and thus offer a possible alternative explanation of the $\Delta \gamma$ measurements in Au+Au collisions without invoking CME interpretation.

\begin{figure}[hbt]
  \begin{center}
    \includegraphics[width=0.45\textwidth]{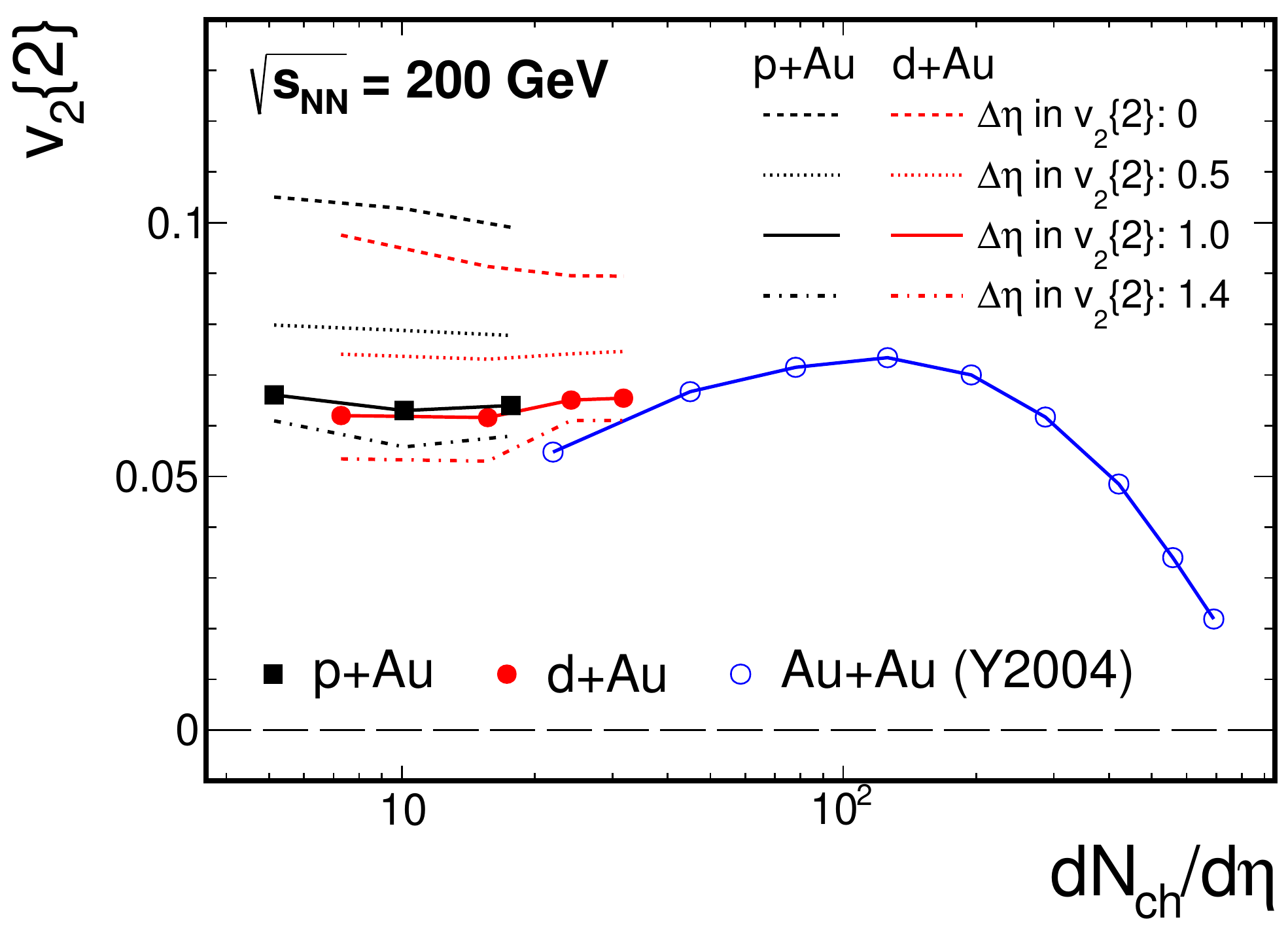}
    \caption{The measured two-particle cumulant $\vtwo$ with $\eta$ gap of 1.0 as a function of multiplicity in \pA\ and \dA\ collisions, compared to that in \AuAu\ collisions~\cite{Abelev:2009ad,Abelev:2009ac}. 
	  The data points connected by solid lines are measured using $\Delta\eta$ gap of 1.0 in $\vtwo$. 
		Results with $\eta$ gaps of 0, 0.5 and 1.4 are shown in dash lines.}
    \label{fig:v2}
  \end{center}
\end{figure}

If indeed dominated by background contributions, 
the $\dg$ may be proportional to the average $v_{2}$ of the background sources, as represented by Eq.~(\ref{eq:bkgd}). 
The $v_{2}$ of the background sources likely scale with the $v_{2}$ of the final-state particles that are measured.
The background should also be proportional to the number of background sources, and because $\dg$ is a pair-wise average,
the background is also inversely proportional to the total number of pairs.
As the number of background sources likely scales with $\dNch$, 
thus $\dg$ approximately scales with $v_2/\dNch$. To gain more insight, a scaled $\dg$ observable is introduced:

\begin{linenomath}
\begin{equation}
	\dgscale=\dg\times \dNch/v_2\,.
	\label{eq:scale}
\end{equation}
\end{linenomath}
Since in our analysis there is no distinction between particles $\alpha, \beta$ and $c$ except the electric charge, the $v_2$ in Eq.~(\ref{eq:scale}) is the same as $\vc$.
Figure ~\ref{fig:v2} shows the measured $v_2$ by the two-particle cumulant method with various $\eta$ gaps as a function of multiplicity in \pA, \dA\ collisions, 
together with results from \AuAu\ ~\cite{Abelev:2009ad,Abelev:2009ac} collisions. 
The results show that $v_{2}\{2\}$ is large in \pA\ and \dA\ collisions, and comparable to \AuAu\ results. 
HIJING~\cite{Wang:1991hta} simulation studies of \pA\ and \dA\ collisions suggest significant contribution of nonflow correlations to $v_{2}$ at very low multiplicities. 
Evidence of contribution to $v_{2}$ from collective flow has also been observed at RHIC and the LHC from long-range particle correlations in small systems, 
especially at higher multiplicity~\cite{Khachatryan:2010gv,CMS:2012qk,Abelev:2012ola,Aad:2012gla,Aidala:2016vgl}.

\begin{figure}[hbt]
  \begin{center}
    \includegraphics[width=0.45\textwidth]{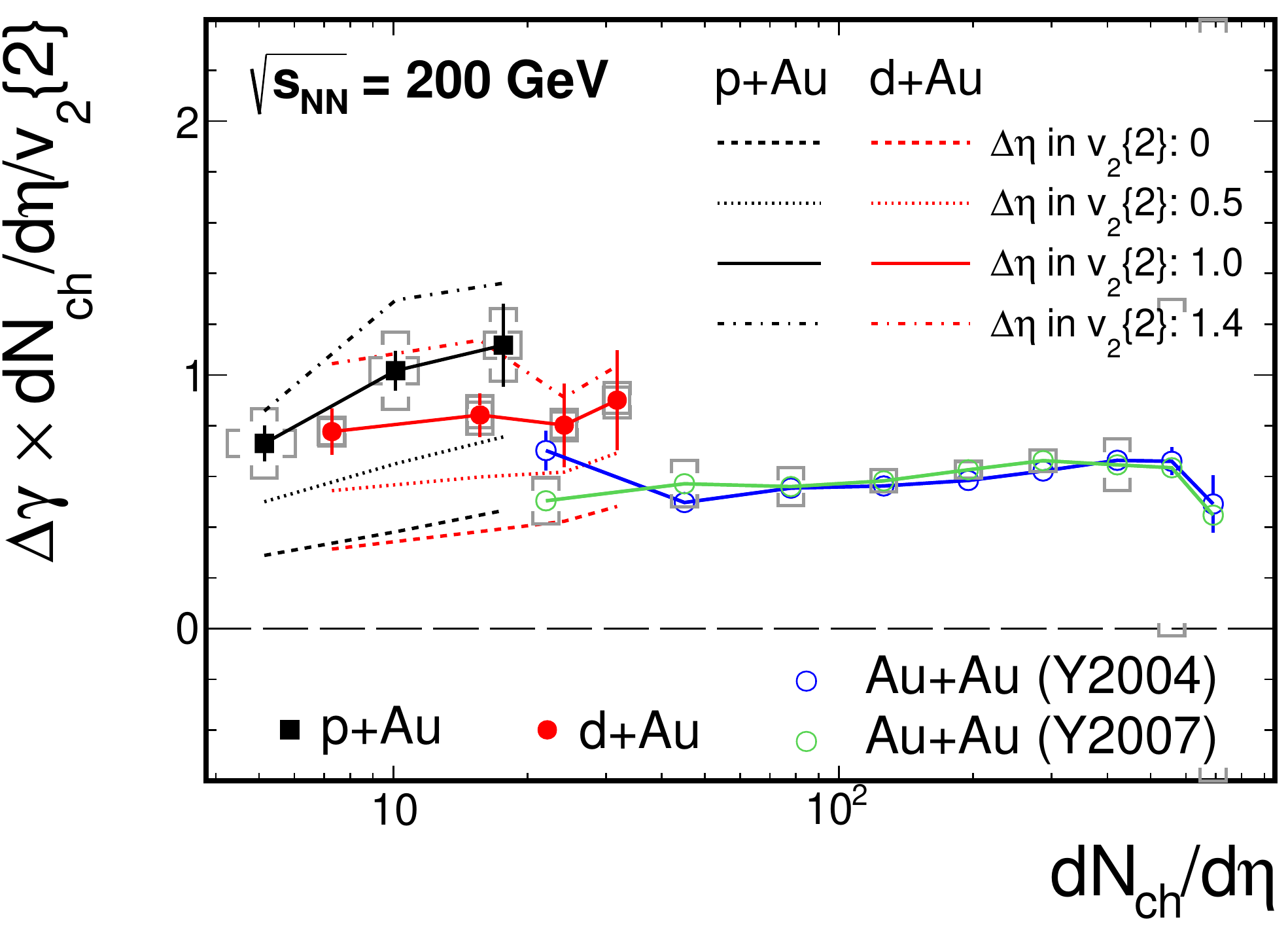}
    \caption{The $\dg\times \dNch/v_2$ in \pA\ and \dA\ collisions as a function of multiplicity, compared to that in \AuAu\ collisions~\cite{Abelev:2009ad,Abelev:2009ac,Adamczyk:2013hsi}. 
	  The data points connected by solid lines are measured using $\Delta\eta$ gap of 1.0 in $\vtwo$. 
	  Dashed lines represent the results using $\vc$ with $\eta$ gaps of 0, 0.5 and 1.4.
	  }	
    \label{fig:scaled}
  \end{center}
\end{figure}

Figure~\ref{fig:scaled} shows the scaled observable $\dgscale$ as a function of multiplicity in \pA\ and \dA\ collisions, 
and compares to that in \AuAu\ collisions. 
Results with different $\eta$ gaps for $\vc$ are also shown. 
The $\dgscale$ in \pA\ and \dA\ collisions are similar to that in \AuAu\ collisions. 
For both small-system and heavy-ion collisions, the $\dgscale$ is approximately constant over $\dNch$,
although within large systematic uncertainties.
Since \pA\ and \dA\ results are dominated by background contributions,
the approximate $\dNch$-independent $\dgscale$ over the wide range of multiplicity in \AuAu\ collisions is consistent with the background scenario. 
Future measurements with larger $\eta$ gaps, especially utilizing upgraded forward detectors,
have the potential to significantly suppress short-range background correlations. 
Those studies will help further to understand the background behavior and differentiate it from the possible CME signal.

\section{Conclusions}
Experimental measurements of $\dg$ in heavy--ion collisions suffer from major backgrounds. 
It is expected that the $\dg$ correlator from small-system \pA\ and \dA\ collisions will be dominated by background correlations, 
as CME-induced contributions would be strongly suppressed due to the random orientations of the magnetic field and the participant plane. 
We reported here measurements of large $\dg$ magnitudes in \pA\ and \dA\ collisions, 
comparable to the values previously reported for peripheral \AuAu\ collisions at similar multiplicities ($\dNch$).
This is similar to the observation at the LHC, where a large $\dg$ signal is observed in $p$+Pb collisions and is comparable to that in Pb+Pb collisions. 
The scaled quantity, $\dg\times \dNch/v_2$, is approximately constant over $\dNch$ for each of the collision systems studied, 
a result expected if background sources dominate.
Our new $p$+Au and $d$+Au measurements, 
where CME contribution is negligible, demonstrate that background contributions could produce magnitudes of the $\dg$ correlator comparable to what has been observed in Au+Au data.
These backgrounds come from particle correlations (such as resonance decays) that are propagated to the $\dg$ observable through correlations to the third particle $c$. 
Our results, while they do not rule out the CME, offer a possible alternative explanation of the $\Delta \gamma$ measurements in  Au+Au collisions without invoking CME interpretation.
New observables~\cite{Magdy:2017yje} and more differential measurements~\cite{Zhao:2017nfq,Xu:2017qfs} are needed to understand the nature of backgrounds and extract any part of the correlations that may be from the CME.
Isobaric collisions taken at RHIC~\cite{Skokov:2016yrj} will further help elucidate the respective CME and background contributions.

\section*{Acknowledgments}

We thank the RHIC Operations Group and RCF at BNL, the NERSC Center at LBNL, and the Open Science Grid consortium for providing resources and support.  This work was supported in part by the Office of Nuclear Physics within the U.S. DOE Office of Science, the U.S. National Science Foundation, the Ministry of Education and Science of the Russian Federation, National Natural Science Foundation of China, Chinese Academy of Science, the Ministry of Science and Technology of China and the Chinese Ministry of Education, the National Research Foundation of Korea, Czech Science Foundation and Ministry of Education, Youth and Sports of the Czech Republic, Hungarian National Research, Development and Innovation Office (FK-123824), New National Excellency Programme of the Hungarian Ministry of Human Capacities (UNKP-18-4), Department of Atomic Energy and Department of Science and Technology of the Government of India, the National Science Centre of Poland, the Ministry  of Science, Education and Sports of the Republic of Croatia, RosAtom of Russia and German Bundesministerium fur Bildung, Wissenschaft, Forschung and Technologie (BMBF) and the Helmholtz Association.

\bibliographystyle{unsrt}
\bibliography{ref}


\end{document}

%% file: star_authors.tex
\affiliation{Abilene Christian University, Abilene, Texas   79699}
\affiliation{AGH University of Science and Technology, FPACS, Cracow 30-059, Poland}
\affiliation{Alikhanov Institute for Theoretical and Experimental Physics, Moscow 117218, Russia}
\affiliation{Argonne National Laboratory, Argonne, Illinois 60439}
\affiliation{American Univerisity of Cairo, Cairo, Egypt}
\affiliation{Brookhaven National Laboratory, Upton, New York 11973}
\affiliation{University of California, Berkeley, California 94720}
\affiliation{University of California, Davis, California 95616}
\affiliation{University of California, Los Angeles, California 90095}
\affiliation{University of California, Riverside, California 92521}
\affiliation{Central China Normal University, Wuhan, Hubei 430079 }
\affiliation{University of Illinois at Chicago, Chicago, Illinois 60607}
\affiliation{Creighton University, Omaha, Nebraska 68178}
\affiliation{Czech Technical University in Prague, FNSPE, Prague 115 19, Czech Republic}
\affiliation{Technische Universit\"at Darmstadt, Darmstadt 64289, Germany}
\affiliation{E\"otv\"os Lor\'and University, Budapest, Hungary H-1117}
\affiliation{Frankfurt Institute for Advanced Studies FIAS, Frankfurt 60438, Germany}
\affiliation{Fudan University, Shanghai, 200433 }
\affiliation{University of Heidelberg, Heidelberg 69120, Germany }
\affiliation{University of Houston, Houston, Texas 77204}
\affiliation{Huzhou University, Huzhou, Zhejiang  313000}
\affiliation{Indian Institute of Science Education and Research (IISER), Berhampur 760010 , India}
\affiliation{Indian Institute of Science Education and Research, Tirupati 517507, India}
\affiliation{Indian Institute Technology, Patna, Bihar, India}
\affiliation{Indiana University, Bloomington, Indiana 47408}
\affiliation{Institute of Physics, Bhubaneswar 751005, India}
\affiliation{University of Jammu, Jammu 180001, India}
\affiliation{Joint Institute for Nuclear Research, Dubna 141 980, Russia}
\affiliation{Kent State University, Kent, Ohio 44242}
\affiliation{University of Kentucky, Lexington, Kentucky 40506-0055}
\affiliation{Lawrence Berkeley National Laboratory, Berkeley, California 94720}
\affiliation{Lehigh University, Bethlehem, Pennsylvania 18015}
\affiliation{Max-Planck-Institut f\"ur Physik, Munich 80805, Germany}
\affiliation{Michigan State University, East Lansing, Michigan 48824}
\affiliation{National Research Nuclear University MEPhI, Moscow 115409, Russia}
\affiliation{National Institute of Science Education and Research, HBNI, Jatni 752050, India}
\affiliation{National Cheng Kung University, Tainan 70101 }
\affiliation{Nuclear Physics Institute of the CAS, Rez 250 68, Czech Republic}
\affiliation{Ohio State University, Columbus, Ohio 43210}
\affiliation{Panjab University, Chandigarh 160014, India}
\affiliation{Pennsylvania State University, University Park, Pennsylvania 16802}
\affiliation{NRC "Kurchatov Institute", Institute of High Energy Physics, Protvino 142281, Russia}
\affiliation{Purdue University, West Lafayette, Indiana 47907}
\affiliation{Pusan National University, Pusan 46241, Korea}
\affiliation{Rice University, Houston, Texas 77251}
\affiliation{Rutgers University, Piscataway, New Jersey 08854}
\affiliation{Universidade de S\~ao Paulo, S\~ao Paulo, Brazil 05314-970}
\affiliation{University of Science and Technology of China, Hefei, Anhui 230026}
\affiliation{Shandong University, Qingdao, Shandong 266237}
\affiliation{Shanghai Institute of Applied Physics, Chinese Academy of Sciences, Shanghai 201800}
\affiliation{Southern Connecticut State University, New Haven, Connecticut 06515}
\affiliation{State University of New York, Stony Brook, New York 11794}
\affiliation{Temple University, Philadelphia, Pennsylvania 19122}
\affiliation{Texas A\&M University, College Station, Texas 77843}
\affiliation{University of Texas, Austin, Texas 78712}
\affiliation{Tsinghua University, Beijing 100084}
\affiliation{University of Tsukuba, Tsukuba, Ibaraki 305-8571, Japan}
\affiliation{United States Naval Academy, Annapolis, Maryland 21402}
\affiliation{Valparaiso University, Valparaiso, Indiana 46383}
\affiliation{Variable Energy Cyclotron Centre, Kolkata 700064, India}
\affiliation{Warsaw University of Technology, Warsaw 00-661, Poland}
\affiliation{Wayne State University, Detroit, Michigan 48201}
\affiliation{Yale University, New Haven, Connecticut 06520}

\author{J.~Adam}\affiliation{Brookhaven National Laboratory, Upton, New York 11973}
\author{L.~Adamczyk}\affiliation{AGH University of Science and Technology, FPACS, Cracow 30-059, Poland}
\author{J.~R.~Adams}\affiliation{Ohio State University, Columbus, Ohio 43210}
\author{J.~K.~Adkins}\affiliation{University of Kentucky, Lexington, Kentucky 40506-0055}
\author{G.~Agakishiev}\affiliation{Joint Institute for Nuclear Research, Dubna 141 980, Russia}
\author{M.~M.~Aggarwal}\affiliation{Panjab University, Chandigarh 160014, India}
\author{Z.~Ahammed}\affiliation{Variable Energy Cyclotron Centre, Kolkata 700064, India}
\author{I.~Alekseev}\affiliation{Alikhanov Institute for Theoretical and Experimental Physics, Moscow 117218, Russia}\affiliation{National Research Nuclear University MEPhI, Moscow 115409, Russia}
\author{D.~M.~Anderson}\affiliation{Texas A\&M University, College Station, Texas 77843}
\author{R.~Aoyama}\affiliation{University of Tsukuba, Tsukuba, Ibaraki 305-8571, Japan}
% ...

\collaboration{STAR Collaboration}\noaffiliation